# Millisecond Electron-Phonon Relaxation in Ultrathin Disordered Metal Films at Millikelvin Temperatures

M. E. Gershenson[a)], D. Gong, and T. Sato

*Department of Physics and Astronomy, Rutgers University, Piscataway, NJ 08854*

B.S. Karasik

*Jet Propulsion Laboratory, California Institute of Technology, Pasadena, CA 91109*

A.V. Sergeev

*Dept. of Electrical and Computer Engineering, Wayne State University, Detroit, MI 48202*

We have measured directly the thermal conductance between electrons and phonons in ultra-thin Hf and Ti films at millikelvin temperatures. The experimental data indicate that electron-phonon coupling in these films is significantly suppressed by disorder. The electron cooling time $\tau_\varepsilon$ follows the $T^{-4}$-dependence with a record-long value $\tau_\varepsilon = 25$ ms at $T = 0.04$K. The hot-electron detectors of far-infrared radiation, fabricated from such films, are expected to have a very high sensitivity. The noise equivalent power of a detector with the area 1 $\mu m^2$ would be $(2-3) \times 10^{-20}$ W$\sqrt{Hz}$, which is two orders of magnitude smaller than that of the state-of-the-art bolometers.

---

a)    Author to whom correspondence should be addressed
      E-mail: gersh@physics.rutgers.edu

Future space far-infrared (FIR) radioastronomy missions will require significant improvement in the sensitivity of radiation detectors in the 40-500 μm wavelength range and integration of the detectors in large arrays for faster sky mapping [1]. The photon-noise-limited noise equivalent power (*NEP*) of a detector combined with a cooled space telescope is expected to be ~ $1 \times 10^{-19}$ W/√Hz [2], or even lower for narrow band applications. The state-of-the-art conventional bolometers currently demonstrate *NEP* ~ $10^{-17}$ W/√Hz at 0.1 K, along with the time constant τ ~ $10^{-3}$ s [3,4].

Recently, we proposed a concept for a submm/FIR direct detector based on disorder-controlled electron heating in superconducting microbridges [5]. The hot-electron direct detectors (HEDDs) operating at millikelvin temperatures can offer unparalleled sensitivity, along with the simplicity of fabrication on bulk substrates, integration with planar antennas, and large-array scalability.

For HEDDs, both decrease of the device volume and increase of the time constant improve the sensitivity [5]. The time constant of HEDDs is determined by the electron cooling time, $\tau_\varepsilon$, due to electron-phonon relaxation. To achieve high sensitivity, the materials with long time constant $\tau_\varepsilon$ are needed. The slowest energy relaxation is expected in properly designed disordered superconducting films [5]. Since the first measurements of the electron-phonon relaxation time in metal films at $T$ = 25-320 mK [6], not much has been added to our knowledge of the electron-phonon processes at millikelvin temperatures. In this Letter, we present measurements of the electron-phonon cooling time at $T$ < 1 K in disordered Hf and Ti thin films, which are promising candidates for the HEDD sensor elements.

We have measured the electron-phonon cooling time in thin films of Hf and Ti on sapphire substrates (the parameters of three samples are listed in Table 1). The films with the critical temperature $T_c$ = 0.3-0.5K and the superconducting transition width $\Delta T_c \approx$ 2-7 mK were deposited by dc magnetron sputtering. They were lithographically patterned into 5 μm wide and 10 cm long strips shaped as meanders. The large length ensures that one can neglect the outdiffusion of hot electrons into cold leads [7] and that the electron-phonon coupling is the only mechanism for electron cooling.

For measuring $\tau_\varepsilon$, we used the well-developed hot-electron technique (see, e.g. [8]). At $T < 1$K, the electron-electron scattering rate exceeds the electron-phonon one by many orders of magnitude [9], and the non-equilibrium distribution function for electrons is well-thermalized. The thermal conductivity between electrons and phonons, $G_{\text{e-ph}} = C_e/\tau_\varepsilon$, can be found from the energy balance equation [8]

$$\frac{P}{V} = \int_T^{T_e} \frac{C_e(T)}{\tau_e(T)} dT = \frac{g\left(T_e^{a+2} - T^{a+2}\right)}{(a+2)\,\tau_e(1K)}, \qquad (1)$$

where $C_e = \gamma T$ is the electron heat capacity, $\tau_\varepsilon = \tau_\varepsilon(1\text{K}) \cdot T^{-\alpha}$, $P = I^2 R$ is the Joule power dissipated in a thin film of volume $V$, $T_e$ is the non-equilibrium electron temperature, and $T$ is the equilibrium phonon (bath) temperature. By applying different amounts of the Joule power to the film at a fixed bath temperature, and measuring the corresponding increase of $T_e$ one can find $\tau_\varepsilon$ from Eq. 1 [for $T_e - T \ll T_e$, Eq. 1 can be linearized as $\tau_e(T) = C_e(T) V (T_e - T)/P$].

In our experiment, the resistance of a sample was measured at a very small ac current, $I_{ac}$, by a resistance bridge as a function of the bath temperature and the heating dc current $I_{dc}$. As the electron "thermometer" above the critical temperature ($T > T_C$), we used the temperature dependence of the quantum corrections to the normal-state resistance [9]. Below $T_C$, the sample was driven into the resistive state by the magnetic field $B$. The resistive state is very sensitive to the electron overheating; this allows to measure $\tau_\varepsilon$ with an unparalleled accuracy.

We have verified that, when the measurements are performed at $T < T_c$ in the *resistive* state, the extracted cooling time does not depend on $B$. Only in this case the non-linear effects in superconductivity, e.g., depairing, can be safely ignored. Comparison between the data obtained in the normal and resistive states is shown in Fig. 1. Firstly, the dependence $\tau_\varepsilon(T)$ was extracted from the hot-electron measurements in the resistive state at different values of $B$, which corresponded to the sample resistance $R \sim (0.2 - 0.9)R_N$, $R_N$ is the resistance in the normal state. Secondly, the resistance was measured as a function of $I_{dc}$ at a fixed $T < T_c$ in the *normal* state, when superconductivity was completely suppressed by the magnetic field. To obtain the dependence $R(T_e)$, shown in Fig. 1, the electron temperature $T_e$ for each value of $I_{dc}$ was calculated from Eq. 1 using the $\tau_\varepsilon$ data obtained from the measurements in the *resistive* state. Figure 1 shows that the dependence $R(T_e)$ coincides with the temperature dependence $R(T)$ measured in equilibrium ($I_{dc} = 0$). The latter dependence is due to the logarithmic quantum corrections to the resistance in a two-dimensional film [9] (dashed line is a guide to the eye). This coincidence rules out non-thermal effects in the resistive state, and allows us to facilitate measurements by taking advantage of a very high sensitivity of the resistive state to electron overheating.

The temperature dependences of the thermal conductivity between electrons and phonons, $G_{e\text{-}ph}$, measured for two Hf samples with different $T_c$ are shown in Fig. 2. With lowering the temperature, $G_{e\text{-}ph}$ decreases rapidly as $T^5$. For comparison, we also plot the theoretical temperature dependence of the thermal conductivity between the film and sapphire substrate [10], $G_b = (R_b d)^{-1} \sim T^3$ ($R_b$ is the thermal boundary resistance between the film and the substrate, $d$ is the film thickness). Figure 2 shows that the lower the temperature, the easier to realize the hot-electron regime, $G_{e\text{-}ph} << G_b$, when the electron-phonon coupling is the bottleneck of the energy transfer from hot electrons to the environment.

In order to calculate the electron cooling time from $G_{e\text{-}ph}$, we used the "bulk" values of the electron heat capacity for Hf and Ti (see Table 1). The temperature dependences of $\tau_\varepsilon$ for these films are shown in Fig. 3. The cooling time exceeds 1 ms at $T \sim 0.1$ K, and becomes record-long, up to 25-30 ms, at the lowest temperatures $T$ = 30-40 mK. Saturation of the temperature dependence of $\tau_\varepsilon$ below 0.1 K, observed for sample 3, can be tentatively attributed to electron overheating by the electronic noise in the experimental set-up. Indeed, the noise power, which is sufficient to overheat the electrons by ~ 10 mK at $T$ = 50 mK, does not exceed ~ $3 \cdot 10^{-14}$ W (the corresponding noise current ~ 0.3 nA) even for our "macroscopic" samples.

The experimental dependence $\tau_\varepsilon(T) \propto T^{-4}$ is consistent with predictions of the theory of electron-phonon interaction in disordered metals for the "dirty" limit, when the electron scatterers (impurities, defects, and film/substrate interface potential) vibrate the same way as the host atoms [11]. The condition of the "dirty" limit, $q_T l \ll 1$ ($q_T$ is the wave number of thermal phonons, $l$ is the electron mean free path), is satisfied in the studied films over a broad range $T <$ 50 K. In disordered films, the transverse phonons govern the electron-phonon coupling [12]. Acoustic impedances of Hf and Ti are close to the impedance of sapphire substrate ($\rho u_t$  $2.6 \times 10^7$ g/m²s), so vibrations of the film-substrate interface are expected to be identical to the phonon modes in the film. Under these conditions, the electron cooling rate is given by [13]

$$\tau_\varepsilon^{-1} = a \tau_{e-ph}^{-1}(T, \varepsilon_F) = a \frac{\sigma^4}{20} \frac{\sigma v_F}{e^2 \hbar^2} \frac{(k_B T)^4}{\rho u_t^5} , \qquad (2)$$

where $\tau_{e-ph}^{-1}(T,\varepsilon_F)$ is the electron-phonon scattering rate for an electron at the Fermi level ($\varepsilon = \varepsilon_F$), $\sigma$ is the film conductivity, $u_t$ is the transverse sound velocity, $v_F$ is the Fermi velocity, $\rho$ is the density of the film. The numerical factor, $a$, describes energy averaging ($a$ = 0.107 for the $T^4$-dependence of the relaxation rate [14]). The calculated dependences $\tau_\varepsilon(T)$ are shown in Fig. 3. For both materials, an excellent agreement between the experimental data and the theory was obtained *with no fitting parameters*.

It is worth mentioning that the dependence $\tau_\varepsilon \propto T^{-4}$ has been previously observed in disordered Bi films in a temperature rage 0.7-3K [15]. Electron energy relaxation in Hf and Ti films is by two orders of magnitude *slower* than that in Bi films at the same temperature. This substantial quantitative difference can be attributed to the difference in sound velocities. According to Eq. 2, the relaxation rate is inversely proportional to $u_t^5$, and for Bi, $u_t \sim 1 \times 10^3$ m/s, which is 2-3 times smaller than that for Ti and Hf.

In conclusion, our experimental results show that the electron-phonon coupling in thin films can be sufficiently suppressed due to disorder and due to the strong temperature dependence of $\tau_\varepsilon$. The hot-electron FIR detectors with the sensor area of the order of 1 $\mu m^2$ would demonstrate a noise equivalent power $NEP = (4k_B T^2 G_{e-ph} V)^{1/2} \approx (2-3) \times 10^{-20}$ W/$\sqrt{Hz}$ at $T = 0.1$ K [16] ($k_B$ is the Boltzman constant), i.e., at least two orders of magnitude better than that for the state-of-the-art bolometers. In a micron-size sensor, the outdiffusion of hot electrons has to be blocked by Andreev reflection from the current leads fabricated from a superconductor with a superconducting energy gap $\Delta$ much larger than that of the sensor [17]. We expect that an antenna- or waveguide-coupled micron-size HEDD with a small time constant $\tau \sim 10^{-3} \div 10^{-5}$ s will exhibit at $T = 0.1$-$0.3$ K the photon-noise-limited performance in millimeter, sub-millimeter, and infrared wavelengths [18].

This research was supported by the Caltech President's Fund, NASA, and ARO MURI.

**References**

1. D. Leisawitz, W. Danchi, M. DiPirro, L.D. Feinberg, D. Gezari, M. Hagopian, W.D. Langer, J.C. Mather, S.H. Mosley, Jr., M. Shao, R.F. Silverberg, J. Staguhn, M.R. Swain, H.W. Yorke, and X. Zhang, *in UV, Optical and IR Space Telescopes and Instruments*, *Proc. SPIE* **4013**, 36 (2000).

2. J. C. Mather, S. H. Moseley, Jr., D. Leisawitz, E. Dwek, P. Hacking, M. Harwit, L. G. Mundy, R. F. Mushotzky, D. Neufeld, D. Spergel, E. L. Wright, *http://xxx.lanl.gov/abs/astro-ph/9812454*

3. J. J. Bock, J. Glenn, S. M. Grannan, K. D. Irwin, A. E. Lange, H. G. LeDuc, and A. D. Turner, *Proc. SPIE* **3357**, 297 (1998).

4. J. M Gildemeister, A. T. Lee, and P. L. Richards, *Appl. Phys. Lett*. **74**, 868 (1999).

5. B. S. Karasik, W. R. McGrath, H. G. LeDuc, M. E. Gershenson, *Supercond. Sci. Technol.* **12**, 745 (1999).

6. M. L. Roukes, M. R. Freeman, R. S. Germain, R. C. Richardson, M. B. Ketchen, *Phys. Rev. Lett*. **55**, 422 (1985).

7. D. Prober, *Appl. Phys. Lett*. **62**, 2119 (1993).

8. G. Bergmann, W. Wei, Y. Zou, and R. M. Mueller, *Phys. Rev*. B **41**, 7386 (1990); P. M. Echternach, M. R. Thoman, C. M. Gould, and H. M. Bozler, *Phys. Rev*. B **46**, 10339 (1992); F. C. Wellstood, C. Urbina, and J. Clarke, *Phys. Rev*. B **49**, 5942 (1994).

9. B. L. Altshuler, A. G. Aronov, M. E. Gershenson, and Yu. V. Sharvin, *Sov. Sci. Rev* A **9**, 223 (1987).

10. E. T. Swartz and R. O. Pöhl, *Rev. Mod. Phys*. **61**, 605 (1989).

11. A. Sergeev and V. Mitin, *Phys. Rev. B.* **61**, 6041 (2000).

12. A. Sergeev and M. Reizer, *Int. J. Mod. Phys.* **10**, 635 (1996).

13. M. Yu. Reizer and A. V. Sergeev, *Sov. Phys.- JETP* **63**, 616 (1986); J. Rammer and A. Schmid, *Phys. Rev.* B **34**, 1352 (1987).

14. K. S. Il'in, N. G. Ptitsina, A. V. Sergeev, G. N. Goltsman, E. M. Gershenzon, B. S. Karasik, E. V. Pechen, and S. I. Krasnosvobodtsev, *Phys. Rev.* B **57**, 15623 (1998).

15. Yu. F. Komnik, V. Yu. Kashirin, B. I. Belevtsev, and E. Yu. Beliaev, *Phys. Rev.* B **50**, 15298 (1994).

16. M. E. Gershenson, D. Gong, T. Sato, B. S. Karasik, W. R. McGrath, and A. V. Sergeev, *Proc. of 11th Int. Symp. on Space Terahertz Technology*, Ann Arbor, MI, 2000, p.514-523.

17. M. Nahum and J. M. Martinis, *Appl. Phys. Lett.* **63**, 3075 (1993).

18. B. S. Karasik, W. R. McGrath, M. E. Gershenson, and A. V. Sergeev, *J. Appl. Phys.*, **87**, 7586 (2000).

19. B. A. Sanborn, P. B. Allen, and D. A. Papaconstantopoulos, *Phys. Rev.* B **40**, 6037 (1989).

**Table 1.** Parameters of the samples

| Sample | Metal | $d$ nm | $T_c$ K | $R_\square(1K)$ Ω | $D$ $10^{-4}$ m²/s | $l$ nm | $v_F$ [19] $10^6$ m/s | $\gamma$ W/(m³K²) | $u_t$ $10^3$ m/s | $\rho$ $10^3$ kg/m³ |
|---|---|---|---|---|---|---|---|---|---|---|
| 1 | Hf | 25 | 0.48 | 38 | 1.48 | 0.94 | 0.47 | 160 | 1.97 | 13 |
| 2 | Hf | 25 | 0.3 | 38 | 1.48 | 0.94 | 0.47 | 160 | 1.97 | 13 |
| 3 | Ti | 20 | 0.43 | 14.7 | 2.44 | 2.3 | 0.32 | 310 | 3.13 | 4.5 |

$R_\square$ is the sheet resistance of the film, $D$ is the electron diffusion constant

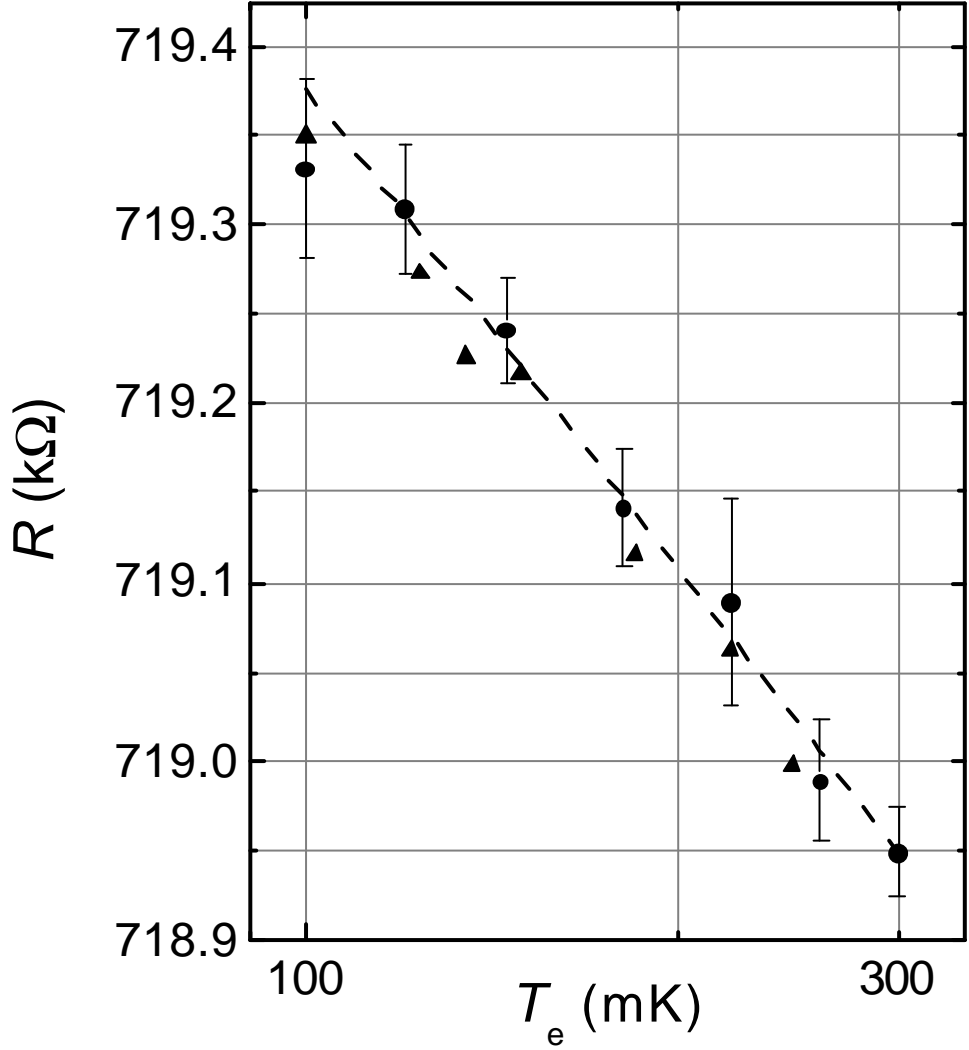

FIG. 1. The dependence $R(T_e)$ (▲) for sample 1, measured at the bath temperature $T = 0.1$ K and $B = 5$T (superconductivity is completely suppressed by the strong magnetic field) for different $I_{dc}$. The electron temperature $T_e(I_{dc})$ was calculated from Eq. 1, where $\tau_\varepsilon$ was obtained from the measurements in the *resistive* state. For comparison, we plot the dependence $R(T_e)$, measured in equilibrium (●, $I_{dc} = 0$, $T_e = T$) at $B = 5$T. In this case, the logarithmic temperature dependence is due to the quantum corrections to the resistance in a two-dimensional film [9] (dashed line is a guide to the eye).

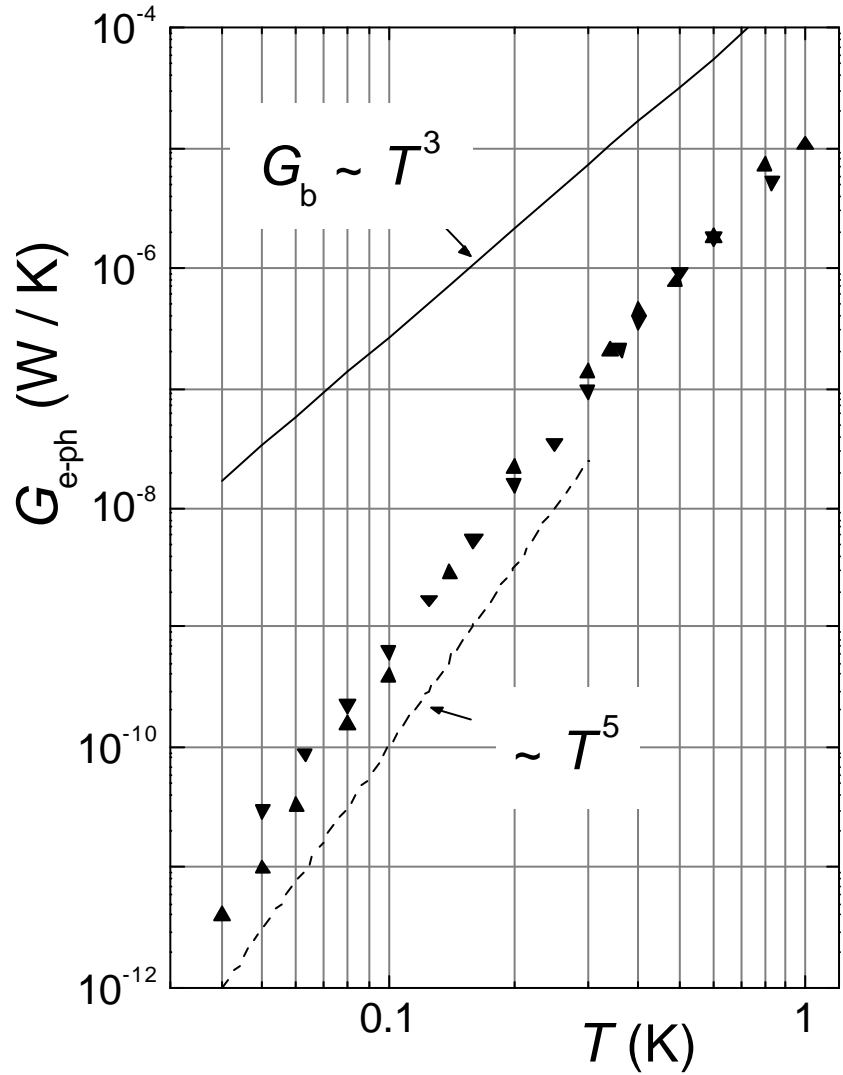

FIG. 2. The temperature dependences of the thermal conductivity $G_{e\text{-}ph}$ for two Hf meanders with the total area $\sim 0.5$ mm$^2$ : ▲ - sample 1, ▼ - sample 2. The solid line represents the thermal conductivity $G_b$ between the meander and the sapphire substrate [10], the dashed line is a guide to the eye.

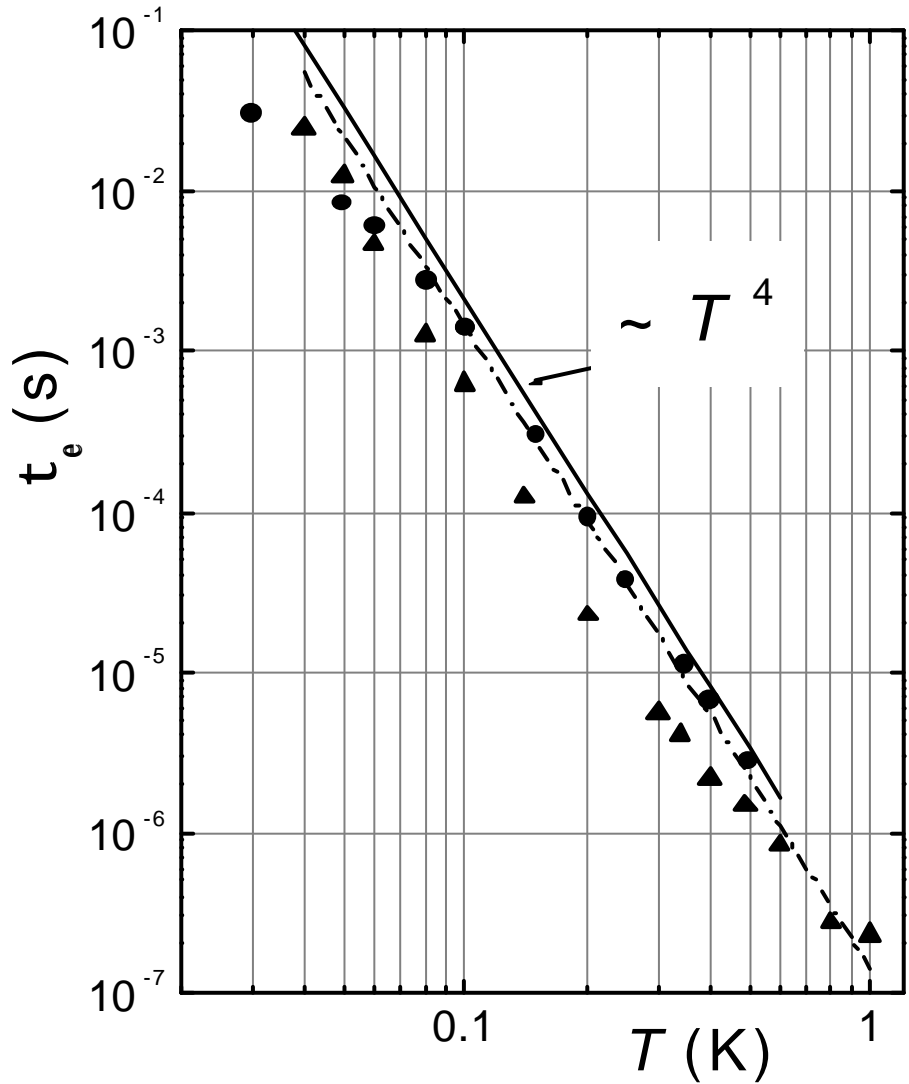

FIG. 3. Temperature dependences of the electron cooling time $\tau_\varepsilon$ for ultra-thin films: ▲ - Hf, sample 1, ● - Ti, sample 3. The lines represent theoretical estimates for $\tau_\varepsilon(T)$ in the "dirty" limit (Eq. 2) (solid line - sample 3, dashed line - sample 1).